\documentclass[showpacs, amsmath, amssymb]{rspublic}
\usepackage[numbers]{natbib} 
\usepackage{color}
\usepackage{amsfonts}
\usepackage{amsbsy}
\usepackage{latexsym}
\usepackage{mathrsfs}
\usepackage{bbm}
\usepackage{bm}
\usepackage[dvipdfm]{graphicx}

\newcommand{\be}{\begin{eqnarray}}
\newcommand{\ee}{\end{eqnarray}}
\def\({\left(}
\def\){\right)}
\def\[{\left[}
\def\]{\right]}

\def\C{\mathbb{ C}}
\def\R{\mathbb{ R}}

\newcommand{\e}{\mathrm{e}}
\newcommand{\im}{\mathrm{i}}

\newcommand{\dd}{\mathrm{d}}

\newcommand{\bra}[1]{\left\langle #1 \right|}
\newcommand{\ket}[1]{\left| #1 \right\rangle}

\newcommand{\ve}[1]{ \boldsymbol #1}

\newcommand{\Tr}{\mathrm{Tr}}
\newcommand{\sla}[1]{\rlap{\kern .15em /}#1}

\newcommand{\ot}{\otimes}

\newcommand{\op}{\oplus}

\begin{document}

\title[Geometric Aspects of Composite Pulses]{Geometric Aspects of Composite Pulses}
\author[T. Ichikawa et al.]{Tsubasa Ichikawa$^{1\footnote{Present address: Department of Physics, Gakushuin University, 1-5-1 Mejiro, Toshima-ku, Tokyo 171-8588, Japan}}$, Masamitsu Bando$^1$, Yasushi Kondo$^{1,2}$ and Mikio Nakahara$^{1,2}$}
\affiliation{$^1$Research Center for Quantum Computing,\\ 
Interdisciplinary Graduate School of Science and Engineering, 
Kinki University, \\
3-4-1 Kowakae, Higashi-Osaka, Osaka 577-8502, Japan\\
$^2$Department of Physics, Kinki University, \\
3-4-1 Kowakae, Higashi-Osaka, Osaka 577-8502, Japan}
\label{firstpage}
 \maketitle

\begin{abstract}{NMR, Composite pulses, Geometric phases, Geometric quantum gates, Quantum control}
Unitary operations acting on a quantum system must be robust 
against systematic errors in control parameters for reliable
quantum computing.
Composite pulse technique in nuclear magnetic 
resonance (NMR) realises such a robust operation 
by employing a sequence of possibly poor quality pulses.
 In this article, we 
demonstrate that two kinds of composite pulses, one compensates for a pulse 
length error in a one-qubit system and the other compensates for a 
$J$-coupling error in
a two-qubit system, have vanishing dynamical phase and thereby can be seen as 
geometric quantum gates, which implement unitary gates by the holonomy 
associated with dynamics of cyclic vectors defined in the text. 
\end{abstract}

\section{Introduction}

Nuclear magnetic resonance (NMR) has developed many techniques to control 
physical systems 
and maintain their coherence \cite{Freeman99, Claridge99}. 
A composite pulse is one of such techniques, in which 
a sequence of pulses is employed to 
cancel out a systematic error inherent in the pulses \cite{Jones09}. 
A systematic error is an unwanted imperfection in control parameters, 
such as poor calibration, and should not be confused with a random noise.
The composite $\pi$-pulse 
by Levitt and Freeman \cite{Levitt79}, developed with intuitive but 
convincing account of its robustness,
opened up a new field of research. 
Now we have hundreds of composite pulses \cite
{Levitt86, Levitt96} and dozens of methods to design them, such as iterative 
expansion \cite{Tycko85}, gradient ascent pulse engineering (GRAPE)
\cite{Khaneja05,Machnes11} and concatenation \cite{Ichikawa11}.

Recently, quantum information processing (QIP)
\cite{Nielsen00, Bengtsson06, Gaitan08, Nakahara08} has an influence over 
the composite pulse design. Very accurate control of a quantum system
is required for a successful quantum error correction, as shown in
\cite{Gaitan08} for example.
Any quantum algorithm can be simulated by quantum circuits composed of
one-qubit unitary operations and the controlled-NOT (CNOT) operations. As a
result, robustness is required for {\it arbitrary} one-qubit operations and 
CNOT operation. In contrast, operations with limited angles and phases
have been required in conventional NMR manipulations.
Numerous composite pulses have been proposed to date in the context of QIP
\cite{Ichikawa11, Jones11, Jones03, Hill07, Testolin07, Tomita10, Cummins03, Brown04, Alway07}.

Geometric quantum computation \cite{Zanardi99, Zhu00} has been
proposed to attain reliable quantum control. In addition to the
dynamical phase, cyclic evolution of a quantum system allows for various
geometric phases \cite{Shapere89, Berry84, Wilczek84, Aharonov87, 
Mead92}, which are controllable and thereby can be utilised for
unitary operations. We call a gate implemented with a geometric phase a
geometric quantum gate (GQG) hereafter. Mathematically, a geometric phase 
is regarded 
as a holonomy associated with a closed path in a suitable base manifold
associated with a cyclic evolution \cite{Simon83, Nakahara03, Page87}. 
Random fluctuations along the integration path are
expected to cancel out, 
leading to a quantum gate robust against random noise. Although there
is numerical support for the robustness of GQGs \cite{Zhu05},
this issue is still under debate \cite{Blais03}.

In this article, we unite these two apparently different constructions of 
robust unitary operations. 
More precisely, we reveal that composite pulses robust against certain kinds of systematic errors are nothing but GQGs. This has been observed previously in 
one-qubit operations \cite{Kondo11}. Now we elaborate and generalise this 
observation to two-qubit operations, which are indispensable for a universal 
set of quantum gates in QIP. Our work
reveals that many composite pulses are
geometric in nature and their robustness is attributed to 
the robustness of GQGs against certain errors.

This article is organised as follows. Geometric phase, in particular
Aharonov-Anandan phase and its application to implementation of
a quantum gate are
introduced in Sec.~2. We employ the perturbation theory as a guiding principle 
to design composite pulses and derive the robustness condition in Sec.~3. 
In Sec.~4, we present the main statement of this article, that is, 
existing composite
pulses to suppress the pulse length error and the $J$-coupling error
are GQGs. We will employ a group theoretical argument to present our
statement in a unified manner.
The assertion in Sec.~4 is exemplified in Sec.~5 and 6 by analysing various 
composite pulses from our viewpoint. Section~7 is 
devoted to conclusion and discussions.

\section{Geometric Quantum Gates}
\label{GQG}

Geometric phase, anticipated in many branches of physics and 
chemistry \cite{Shapere89}, 
was formulated first by Berry in an adiabatic evolution of a quantum system. 
In \cite{Berry84}, Berry considered a cyclic evolution of a quantum system 
whose Hamiltonian has time-dependent parameters, and pointed out that
after the cyclic and adiabatic evolution, the system may acquire not only the 
dynamical phase factor, but also a geometric phase factor, which is given by
a circuit integral in the parameter manifold. This integral is geometric, in the sense
that it is independent of how fast the circuit is traversed.
The Berry phase has been generalised 
in many ways. One of such generalizations is Wilczek-Zee holonomy: 
In the presence of $n$-fold degeneracy, the geometric phase factor can be
replaced to an element of a unitary group U($n$), which is also independent of
how fast the circuit is traversed \cite{Wilczek84}.

Aharonov and Anandan showed in \cite{Aharonov87} that the geometric phase 
appears even in a non-adiabatic evolution. Consider an $n$-level system, 
whose normalised state vector at time $t\in [0,T]$ is given by $\ket{\psi(t)}\in\C^n$. 
Dynamics of the system is characterised by the Schr\"odinger equation
\be
\ri\frac{\rd}{\rd t}|\psi(t)\rangle=H(\lambda(t))\ket{\psi(t)},
\ee
where the Hamiltonian $H(\lambda(t))$ is Hermite and time-dependent through parameters 
$\lambda(t)=(\lambda_1(t), \ldots, \lambda_N(t))$. Here we set $\hbar=1$. 
When the evolution is cyclic with a period $T$, i.e.,
\be
\ket{\psi(T)}=\re^{\ri\gamma}\ket{\psi(0)},
\qquad
\gamma\in\R,
\ee
then the phase $\gamma$ the system acquires after the cyclic
evolution includes geometric contribution $\gamma_{\rm g}$,
which is defined in terms of the dynamical phase $\gamma_{\rm d}$ as
\be
\gamma_{\rm g}=\gamma-\gamma_{\rm d},
\qquad
\gamma_{\rm d}=-\int_0^T\dd t\bra{\psi(t)}H(\lambda(t))\ket{\psi(t)}.
\ee
This phase $\gamma_{\rm g}$ is called the Aharonov-Anandan phase. It is possible to
interpret the Aharonov-Anandan phase in terms of geometric structure of the Hilbert space $\C^n$.
See \ref{appHol} for details.
Also, for another expression of the Aharonov-Anandan phase, see, {\it e.g.}, \cite{Aharonov87, Mead92, Page87}.

Applications of geometric phases are found 
in QIP. For example, Zanardi and Rasetti proposed to use the
Wilczek-Zee holonomy to implement unitary gates \cite{Zanardi99}.
It is also possible to implement unitary gates by using the 
Aharonov-Anandan phase 
\cite{Ichikawa11, Zhu00, Kondo11, Ota09a, Ota09b}. To see this, let 
$\{\ket{\psi_a}\}_{1 \leq a \leq n}$ be the eigenvectors of a
Hamiltonian $H(\lambda(0))$ and suppose their dynamical evolution is cyclic, 
that is,
\be
\ket{\psi_a(T)}=U(T)\ket{\psi_a},
\qquad
U(T)={\cal T}\re^{-\ri\int_0^T\dd s H(\lambda(s))}
\label{adb}
\ee
and
\be
\ket{\psi_a(T)}=\re^{\ri\gamma^a}\ket{\psi_a},
\qquad
\gamma^a\in\R,
\label{cyc}
\ee
where the time-ordered product is denoted by ${\cal T}$.
Equating Eqs.~(\ref{cyc}) and (\ref{adb}), we observe that $\ket{\psi_a}$
is an eigenvector of $U(T)$ with the eigenvalue $\re^{\ri\gamma^a}$, 
that is,
\be
U(T)\ket{\psi_a}=\re^{\ri\gamma^a}\ket{\psi_a}.
\label{eigen}
\ee 
When there is no degeneracy, the spectral decomposition of $U(T)$ is written as
\be
U(T)=\re^{\ri\gamma^1}\ket{\psi_1}\bra{\psi_1}+\dots+\re^{\ri\gamma^n}\ket{\psi_n}\bra{\psi_n}.
\ee
The phase $\gamma^a$ is decomposed as $\gamma^a=\gamma_{\rm g}^a+
\gamma_{\rm d}^a$ in terms of the dynamical phase defined as
\be
\gamma_{\rm d}^a=-\int_0^T\dd t\bra{\psi_a(t)}H(\lambda(t))\ket{\psi_a(t)},
\qquad
\ket{\psi_a(t)}=U(t)\ket{\psi_a}.
\ee
A unitary operator $U(T)$ is called a geometric quantum 
gate (GQG) if $\gamma_{\rm d}^a$ vanishes for all $a$.

\section{Perturbative Construction of Composite Pulses}

In actual situations in NMR, the dynamics is controlled by a
sequential application of rf-pulses with 
constant field strength. Accordingly, the time interval $[0, T]$ is divided 
into $k$ intervals, in each of which the Hamiltonian is constant. 
More precisely, we define the $i$-th temporal interval $[t_
{i-1}, t_i]$, where $t_i$ satisfies $0=t_0<t_1<\ldots<t_k=T$, and define
a piecewise constant Hamiltonian, which takes the form $H(\lambda^i)$ in 
the $i$-th interval $[t_{i-1}, t_i]$. Here $\lambda^i=(\lambda_1^i,
\dots, \lambda_N^i)$ is a constant parameter vector while $N$ is the
number of control parameters. Then, the $i$-th rf-pulse gives rise 
to a unitary operator
\be
\re^{-\ri W^i},
\qquad
W^i=H(\lambda^i)\cdot(t_i-t_{i-1}),
\ee
and $U(T)$ can be written as
\be
U(T)=\re^{-\ri W^k}\cdots \re^{-\ri W^1}.
\ee
Now we wish to implement a \lq target\rq~unitary operator $U$ as $U=U(T)$. The target $ U$ should be implemented in a way robust against the error under 
consideration as much as possible. 
Hereafter we seek a condition for such robust implementation. 

We consider errors which cause displacement
\be
W^i\rightarrow W^i+\delta W^i,
\label{dis}
\ee
where $\delta W^i$ is a self-adjoint operator corresponding to the error. 
When $\delta W^i$ is sufficiently small 
in the sense of the operator norm, we can use the perturbation theory and
find 
\be
\re^{-\ri (W^i+\delta W^i)}\approx\re^{-\ri W^i}\(\mathbbm{1}_n-\ri\delta W_{\rm I}^i\);
\qquad
\delta W^i_{\rm I}:=\int_0^1\dd x\,\e^{\im x W^i}\delta W^i\e^{-\im x W^i},
\label{discHint}
\ee
to the first order in $\delta W^i$. 
Here the identity operator on $\C^n$ is denoted by $\mathbbm{1}_n$. 
The operator $\delta W^i_{\rm I}$ is the error operator $\delta W^i$ in the interaction picture.
Then, the unitary operator $U^\prime$ implemented with the error $\delta
W^i$ is given by
\be
U^\prime=\re^{-\ri (W^k+\delta W^k)}\cdots \re^{-\ri (W^1+\delta W^1)}\approx U\(\mathbbm{1}
_n-\ri\Delta W\),
\ee
where
\be
\Delta W=\sum_{i=1}^k{V^{i-1}}^\dag\delta W^i_{\rm I}V^{i-1},
\qquad
V^i=\re^{-\ri W^i}\cdots\re^{-\ri W^1}
\quad
{\rm for}
\quad
i=1,2,\dots k-1,
\ee
with $V^0=\mathbbm{1}$. Many, albeit not all, composite pulses satisfy the following robustness
condition
\be
\Delta W=0,
\label{lt}
\ee
which we can evaluate once we specify $\delta W^i$. 
This condition guarantees the effect of the error vanishes to the first order in $\delta W^i$.

Now we wish to address the relation between the robustness condition (\ref{lt}) and a classification of composite pulses common in the NMR community. 
There are two types, Type A and Type B,
of composite pulses \cite{Levitt86, Jones11}. 
The error tolerance is independent of the initial state vector for 
Type~A composite pulses, whereas it is not the case for Type~B 
composite pulses. In view of this, the composite pulses satisfying (\ref{lt}) 
are clearly of Type~A.

\section{Composite Pulses as Geometric Quantum Gates}
\label{CPGQG}
To see the geometric nature of Type A composite pulses, we follow the argument
introduced in \cite{Kondo11}, which has been generalised to multi-qubit system
in \cite{Ichikawa11}. Suppose that the systematic error is proportional to 
$W^i$:
\be
\delta W^i=\epsilon W^i.
\label{eW}
\ee
As shown later, two kinds of systematic errors are of this form.
The robustness condition (\ref{lt}) reads
\be
\Delta W=\epsilon\sum_{i=1}^k{V^{i-1}}^\dag W^iV^{i-1}=0,
\label{DHcom}
\ee
where use has been made of the identity $\delta W^i_{\rm I}=\delta W^i$ derived from Eq.~(\ref{discHint}) and Eq.~(\ref{eW}).
Taking the expectation value of $\Delta W$ with respect to $\ket{\psi_a}$, we 
obtain
\be
\gamma_{\rm d}^a=\sum_{i=1}^k\gamma_{\rm d}^a(i)=0,
\qquad
\gamma_{\rm d}^a(i):=-\bra{\psi_a(i-1)} W^i\ket{\psi_a(i-1)},
\label{nulldyn}
\ee
where $\ket{\psi_a(i)}:=V^{i}\ket{\psi_a}$.
Hence, any composite pulse which is designed by the perturbation theory and compensates 
the error (\ref{eW}) is GQG.
In what follows, we will show that composite pulses associated with two kinds of 
relevant systematic errors are GQGs.

\subsection{Error on One-Qubit System}
We turn to a one-qubit system, whose Hilbert space is $\C^2$. An
SU(2) operations we can 
implement with a single rf-pulse in NMR is limited to the form
\be
W^i=\theta_i\,\bm{n}_i\cdot\bm{\sigma}/2,
\label{decW1}
\ee
where
$
\bm{n}_i=\(\cos\phi_i, \sin\phi_i, 0\)
$ 
and
$\bm{\sigma}=(\sigma_x,\sigma_y,\sigma_z)$ due to the apparatus limitation. 
Nevertheless, we can implement any SU(2) operation by combining at most
three such pulses using the Euler angle decomposition
\cite{Nielsen00, Nakahara08}.
The displacement (\ref{dis}) under the error (\ref{eW}) is seen as
\be
\theta_i\rightarrow(1+\epsilon)\theta_i.
\label{ple}
\ee
This is a well-known systematic error called the pulse length error in the 
NMR community \cite{Levitt86}. Hence, from the previous argument, we observe that any 
composite pulse compensating for the pulse length error is a GQG.

\subsection{Error in Two-Qubit System}
\label{2qubit}
For a two-qubit system, the relevant Hilbert space and the set of unitary
operations are $\C^{2\ot2}$ and ${\rm SU(4)}$, respectively. 
In view of quantum information processing, the
controlled-NOT (CNOT) operation 
\be
U_{\rm CNOT}=\ket{0}\bra{0}\ot\mathbbm{1}_2+\ket{1}\bra{1}\ot\sigma_x
\ee
is important.\footnote{Precisely speaking, $\det U_{\rm CNOT}= -1$ and it is 
not an element of SU(4). Nevertheless, we can multiply this matrix by an
unphysical phase $\re^{i \pi/4}$ to make it an element of SU(4). Two quantum 
gates that differ by an overall phase will be identified hereafter.} 
Here, $\ket{a}\in\C^2$ with $a=0,1$ is the eigenvector of $\sigma_z$ 
with the eigenvalue $(-1)^a$. The relevance of CNOT operation originates from 
the fact that any QIP can be implemented as a quantum circuit composed of
one-qubit unitary operations and CNOT operations 
\cite{Barenco, Nielsen00, Nakahara08}.

By using the Cartan decomposition \cite{Nakahara08}, CNOT operation can be rewritten as
$
U_{\rm CNOT}=K_1H K_2, 
$ 
with
\be
H=\re^{\ri \alpha_x \sigma_x \otimes \sigma_x}
\re^{\ri \alpha_y \sigma_y \otimes \sigma_y}
\re^{\ri \alpha_z \sigma_z \otimes \sigma_z}
\qquad
K_1, K_2\in {\rm SU(2)}\ot{\rm SU(2)}.
\ee
Since 
$\sigma_x \otimes \sigma_x$ is generated from $\sigma_z \otimes \sigma_z$
through the following identity
\be
\re^{\ri \alpha_x \sigma_x\otimes\sigma_x} 
= \re^{\ri\pi(\sigma_y\otimes \mathbbm{1}_2+\mathbbm{1}_2 \ot \sigma_y)/4}
\re^{\ri\alpha_x \sigma_z\otimes \sigma_z}
\re^{-\ri\pi(\sigma_y\otimes \mathbbm{1}_2+\mathbbm{1}_2 \otimes \sigma_y)/4},
\ee 
the Ising-type Hamiltonian
\be
H=J\sigma_z\ot\sigma_z/4
\ee
is essential to implement CNOT operations which is commonly realised in a weak coupling limit.
Hereafter we shall be concerned with the $J$-coupling error defined by
\be
J\rightarrow(1+\epsilon)J.
\label{eO}
\ee

Several composite pulses robust against the $J$-coupling error have been 
proposed assuming that one-qubit operations are free from errors. 
These existing composite 
pulses \cite{Jones03, Hill07, Testolin07, Tomita10} are designed by making use of the following 
three generators only:
\be
X:=\sigma_z\ot\sigma_z,
\qquad
Y:=\sigma_z\ot\sigma_x,
\qquad
Z:=\mathbbm{1}_2\ot\sigma_y, 
\label{xyz}
\ee
among the fifteen generators of SU(4).
Evidently these operators satisfy $\mathfrak{su}(2)$ algebra:
\be
[X/2,Y/2]=\ri Z/2,
\qquad
[Y/2, Z/2]=\ri X/2,
\qquad
[Z/2, X/2]=\ri Y/2.
\ee
Thus, we can construct an SU(2) subgroup by exponentiating the generators 
(\ref{xyz}). Let us denote this subgroup by $G$. 

Now, let us put
\be
&&\Omega_i=J(t_i-t_{i-1})/2,\nonumber\\
&&W^i=\Omega_i\(\cos\phi_i X+\sin\phi_i Y\)/2=\re^{-\ri\phi_iZ/2}\(\Omega_i X/2\)\re^{\ri
\phi_iZ/2}.
\ee
Then, we observe that
\be
\re^{-\im W^i}=\re^{-\ri\phi_iZ/2}\re^{-\im \Omega_iX/2}\re^{\ri\phi_iZ/2}
\in G.
\label{decW2}
\ee
Thus, for this $W^i$, we observe the identification
between $\Omega_i, X, Y$, and $Z$ and 
$\theta_i, \sigma_x, \sigma_y$, and $\sigma_z$ in Eq.~(\ref{decW1}),
respectively. Since the $J$-coupling error (\ref{eO}) is equivalent to 
the pulse 
length error (\ref{ple}) under this identification, we can construct a ``composite pulse'', which is robust against the $J$-coupling error, if we merely replace $\theta_i, \sigma_x, \sigma_y$, and $\sigma_z$ by 
$\Omega_i, X, Y$, and $Z$, respectively. In fact, as 
stated before, such composite pulses based on the identification have been 
proposed in \cite{Jones03, Hill07, Testolin07, Tomita10}. One of these 
composite pulses shall be examined later. Composite pulses designed under
this identification are GQGs, since this identification keeps the mathematical 
structure of the theory unchanged.

Two remarks are in order. First, the definition of $Z$ tells us that we can 
freely tune the parameter $\phi_i$ by changing the rf-field along the $y$-axis
of the second qubit. Second, we can define the Bloch 
sphere for an orbit generated by $G$ and $\ket{\psi}\in\C^{2\ot2}$ 
if $\ket{\psi}$ is an eigenvector of some element $U\in G$. 
In other words, if there exists $U \in G$ such that
\be
U\ket{\psi}=\re^{\ri\gamma}\ket{\psi},
\ee
then the $G$-orbit $G|\psi \rangle$ of $|\psi \rangle$
is identified as the Bloch sphere $S^2$.
This observation ensures that we can visualise the time evolution of a cyclic state associated with $U\in G$ as a trajectory in the Bloch sphere, as long as 
we use the composite pulses proposed so far.

\section{Examples of Geometric Composite Pulse}
In this section, we give several examples demonstrating our claim that
two types of composite pulses introduced in the previous section are GQGs.
To this end, we shall evaluate the
dynamical phase of several composite pulses and verify that the dynamical
phase indeed vanishes in all cases.

\subsection{One-Qubit System}
We parametrise our target $U$ as
\be
U=\exp\(-\ri\theta\bm{n}\cdot\bm{\sigma}/2\),
\qquad
\bm{n}=(\cos\phi, \sin\phi,0).
\label{target}
\ee
Then, from Sec.~\ref{GQG}, a cyclic state $\ket{\psi_a}$ associated with $U$ 
is given as an eigenvector of $U$, that is,
\be
\ket{\psi_a}=\ket{(-1)^{a}\bm{n}},
\qquad
a=0, 1,
\label{bloch_vector}
\ee
where $\ket{(-1)^{a}\bm{n}}$ is the eigenstate of 
$\bm{n}\cdot\bm{\sigma}$ such that
\be
\bm{n}\cdot\bm{\sigma}\ket{(-1)^{a}\bm{n}} = (-1)^{a}\ket{(-1)^{a}\bm{n}}.
\ee
We shall often use the following useful formula:
\be
\bra{\ve{n}}\ve{m}\cdot\ve{\sigma}\ket{\ve{n}}=\ve{n}\cdot\ve{m}.
\label{formula}
\ee
Note that the vector $\bm{n}$ is the Bloch vector for the state $\ket{\bm{n}}$ and
we have
\be
U\ket{(-1)^a\bm{n}}=\omega_a\ket{(-1)^a\bm{n}},
\qquad
\omega_a=\exp[(-1)^{a+1}\ri\theta/2].
\label{eigenU}
\ee

All composite pulses, for which we evaluate the dynamical phases, are
composed from $k=2l-1$ pulses, which satisfy 
the \lq time-symmetric\rq~condition
\be
W^i=W^{k+1-i}.
\ee
Many implications of this condition are found in \cite{Levitt86}. 
Now we address that this condition leads to
\be
\gamma_{\rm d}^a(i) =
\gamma_{\rm d}^a(k+1-i).
\label{gg}
\ee
See \ref{appA} for the proof. Hence, the dynamical phase is rewritten as
\be
\gamma_{\rm d}^a=2\[\gamma_{\rm d}^a(1)+\dots+\gamma_{\rm d}^a(l-1)\]+\gamma_{\rm d}^a(l)
\ee
for a composite pulse, which is made of $k=2l-1$ pulses.

\subsubsection{$90^\circ$-$180^\circ$-$90^\circ$ pulse}
The first composite pulse was proposed by Levitt and Freeman in 1979 based on
a trajectory on the Bloch sphere 
\cite{Levitt79}. This is a $k=3$ symmetric composite pulse defined by
\be
\theta_1=\theta_2/2=\theta_3=\pi/2,
\qquad
\phi_1=\phi_3=0,
\quad
\phi_2=\pi/2.
\ee
We immediately find
\be
W_1=W_3=(\pi/4)\hat{\bm x}\cdot\bm{\sigma},
\qquad
W_2=(\pi/2)\hat{\bm y}\cdot\bm{\sigma},
\ee
which leads to
\be
U=\re^{-\ri W_1}\re^{-\ri W_2}\re^{-\ri W_1}=-\ri\sigma_y.
\label{Uy}
\ee
Hence, we observe that the target is fixed to $\theta=\pi$ and $\phi=\pi/2$ and there are no 
free parameters we may adjust. It follows from Eq.~(\ref{Uy}) that $\ket{\psi_a}=\ket{(-1)^a\hat{\bm{y}}}$.

Let us proceed to the calculation of the dynamical phase. First, we have
\be
\gamma_{\rm d}^a(1)=-(\pi/4)\bra{(-1)^a\hat{\bm{y}}}\hat{\bm x}\cdot\bm{\sigma}\ket{(-1)^a\hat{\bm{y}}}=(-1)^{a+1}(\pi/4)\hat{\bm x}\cdot\hat{\bm{y}}=0
\ee
from the formula (\ref{formula}). Next, we observe
\be
\ket{\psi_a(1)}=\re^{-\ri W^1}\ket{\psi_a}=\re^{-\ri\pi\sigma_x/4}\ket{(-1)^a\hat{\bm y}}=\ket{(-1)^a\hat{\bm z}}
\ee
to obtain
\be
\gamma_{\rm d}^a(2)=-(\pi/2)\bra{(-1)^a\hat{\bm{z}}}\hat{\bm y}\cdot\bm{\sigma}\ket{(-1)^a\hat{\bm{z}}}=(-1)^{a+1}(\pi/2)\hat{\bm y}\cdot\hat{\bm{z}}=0.
\ee
Summing up these, we reach
\be
\gamma_{\rm d}^a=2\gamma_{\rm d}^a(1)+\gamma_{\rm d}^a(2)=0.
\ee

\begin{figure}[t]
\begin{center}
  \includegraphics[width=2.5in]{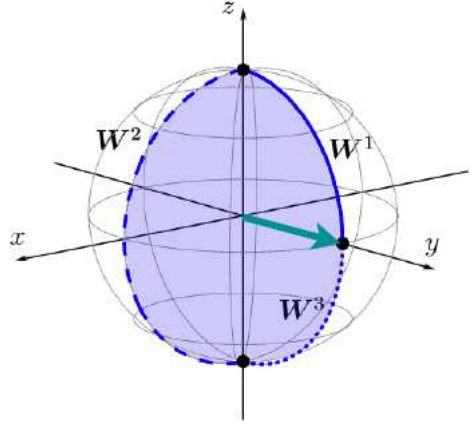}
\caption{(Colour online) Excursion of the cyclic state $\ket{\hat{\bm{y}}}$ under the 90$^\circ$-180$^\circ$-90$^\circ$
pulse on the Bloch sphere. The green arrow is the Bloch vector of the cyclic state. The solid angle of the blue area enclosed by the trajectory of the Bloch vector is equal to $\pi=\theta$, which also shows that this composite pulse is a GQG (See Appendix A).}
\label{levitt}
\end{center}
\end{figure}

We can confirm Eq.~(\ref{gg}) by further calculation. We find
\be
\ket{\psi_a(2)}=\re^{-\ri W^2}\re^{-\ri W^1}\ket{\psi_a}=\re^{-\ri W^2}\ket{(-1)^a\hat{\bm z}}=\ket{(-1)^{a+1}\hat{\bm z}},
\ee
from which it follows that
\be
\gamma_{\rm d}^a(3)=-(\pi/4)\bra{(-1)^{a+1}\hat{\bm{z}}}\hat{\bm x}\cdot\bm{\sigma}\ket{(-1)^{a+1}\hat{\bm{z}}}=(-1)^{a}(\pi/4)\hat{\bm x}\cdot\hat{\bm{z}}=0=\gamma_{\rm d}^a(1).\nonumber\\
\ee
The time-evolution of the cyclic states ends up with
\be
\ket{\psi_a(3)}=\re^{-\ri W^3}\ket{\psi_a(2)}=\ket{(-1)^a\hat{\bm y}}=\ket{\psi_a},
\ee
as expected.
These results are summarised as
\be
\ket{\pm\hat{\bm y}}\xrightarrow{\re^{-\ri W^1}}
\ket{\pm\hat{\bm z}}\xrightarrow{\re^{-\ri W^2}}
\ket{\mp\hat{\bm z}}\xrightarrow{\re^{-\ri W^3}}
\ket{\pm\hat{\bm y}},
\qquad
\gamma_{\rm d}^a(i)=0.
\ee
See Fig.~\ref{levitt} for the graphical representation of this excursion.

The lesson we learn from this composite pulse is that the 
converse of our statement is not always true:
Not all GQGs for a spin-1/2 system are Type A composite 
pulses robust against the pulse length error. Indeed, this pulse is of Type B
since $\Delta W\neq0$. This has been overlooked in \cite{Kondo11}.

\subsubsection{SCROFULOUS}
SCROFULOUS is a $k=3$ time-symmetric composite pulse constructed by Cummins, Llewellyn and Jones \cite{Cummins03}.
This composite pulse was designed by using perturbation theory and quaternion algebra.
Given a target (\ref{target}), SCROFULOUS takes the form
\be
&& \theta_1 = \theta_3 = {\rm arcsinc}[2\cos(\theta/2)/\pi],\qquad \theta_2 = \pi\nonumber\\
&& \phi_1 = \phi_3 = \phi + \arccos[-\pi\cos\theta_1/(2\theta_1\sin(\theta/2))],\nonumber\\
&& \phi_2 = \phi_1 - \arccos[-\pi/(2\theta_1)],
\ee
where ${\rm sinc}\, x=\sin x/x$.
Note that SCROFULOUS implements any one-qubit unitary operator of the form (\ref{target}).

\begin{figure}[t]
\begin{center}
  \includegraphics[width=2.5in]{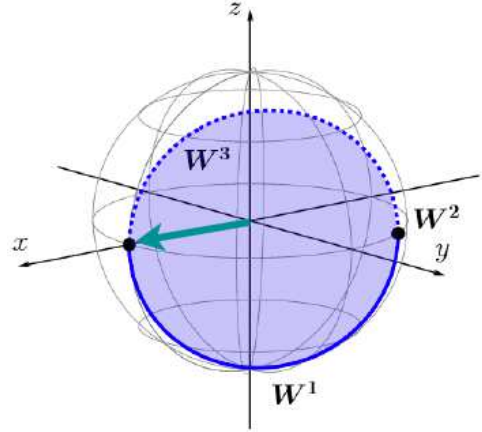}
\caption{(Colour online) Excursion of the cyclic state $\ket{\hat{\bm{x}}}$ of the SCROFULOUS for a target $\theta=\pi, \phi=0$ on the Bloch sphere. The green arrow is the Bloch vector of the cyclic state. The state $\re^{-\ri W^1}\ket{\hat{\bm{x}}}$ pauses during the application of the pulse $W^2$, since it is an eigenstate of the pulse $W^2$. The solid angle of the blue area is equal to $\theta$; this composite pulse is a GQG.}
\label{SCRF}
\end{center}
\end{figure}

Let us evaluate the dynamical phase. We set $\phi=0$ for simplicity,
while extension to an arbitrary $\phi$ is straightforward. First, we have
\begin{equation}
\gamma_{\rm d}^a(1)
= -\theta_1\bra{(-1)^a\bm{n}}\bm{n}_1\cdot\bm{\sigma}/2\ket{(-1)^a\bm{n}}
  = (-1)^{a+1}\theta_1\bm{n}\cdot\bm{n}_1/2= (-1)^{a+1}(\theta_1/2)\cos{\phi_1}.
\end{equation}
Next, observe that
\begin{equation}
{V^1}^\dag W^2V^1
=\frac{\theta_2}{2}
\[
\cos^2(\theta_1/2)\bm{n}_2
+\sin^2(\theta_1/2)\bm{m}
-\sin\theta_1(\bm{n}_1\times\bm{n}_2)
\]
\cdot\bm{\sigma},
\label{lemma1}
\end{equation}
where
\be
\bm{m}=2(\bm{n}_2\cdot\bm{n}_1)\bm{n}_1-\bm{n}_2.
\ee
Since we have
\be
&&\bm{n}_2\cdot\bm{n}=\cos\phi_2=-\pi\cos\phi_1/(2\theta_1)+\sin\phi_1\sin[\arccos(-\pi/(2\theta_1))],\nonumber\\
&&\bm{m}\cdot\bm{n}=\cos(2\phi_1-\phi_2)=-\pi\cos\phi_1/(2\theta_1)-\sin\phi_1\sin[\arccos(-\pi/(2\theta_1))],\nonumber\\
&&(\bm{n}_1\times\bm{n}_2)\cdot\bm{n}=0,
\ee
we observe
\be
&&\gamma_{\rm d}^a(2)=-\bra{\psi_a}V^{1\dag}W^2V^1\ket{\psi_a}\nonumber\\
&&\hspace{26pt}=(-1)^a(\pi/2)\{(\pi/(2\theta_1))\cos\phi_1 + \cos\theta_1\sin[\arccos(-\pi/
(2\theta_1))]\sin\phi_1\}.\nonumber\\
\ee
Using $\sin(\arccos x) = \sqrt{1-x^2}$ and
\begin{eqnarray}
 \sin\phi_1 
 = \sqrt{1-(\pi/(4\theta_1))^2}/\sin(\theta/2),
\end{eqnarray}
we immediately derive
\be
 &&\gamma_{\rm d}^a = 2\gamma_{\rm d}^a(1)+\gamma_{\rm d}^a(2)\nonumber\\
 &&\quad\ = (-1)^a\{\theta_1[1-(\pi/(2\theta_1))^2]\cos\phi_1
   + (\pi/2)\cos\theta_1\sin[\arccos(-\pi/(2\theta_1))]\sin\phi_1\}\nonumber\\
 &&\quad\ = (-1)^a[1-(\pi/(2\theta_1))^2]
  [\theta_1\cos\phi_1 + \pi\cos\theta_1/(2\sin(\theta/2))]\nonumber\\
 &&\quad\ = 0.
\ee
Hence SCROFULOUS is a GQG.
The trajectory of the cyclic state is given in Fig.~\ref{SCRF}.

\subsubsection{Broad Band 1 (BB1)}
Now we turn to the BB1, which was proposed by Wimperis \cite{Wimperis94}.
For brevity\rq s sake, we treat a $k=5$ time-symmetric variant of the BB1 sequence. 
We call this variant time-symmetric BB1. The BB1 pulse sequence is useful 
for the implementation of QIP, since it compensates for the pulse length 
error up to the second order in perturbative expansion \cite{Jones11}. 
There are two techniques to generalise the BB1 pulse sequence \cite{Brown04}. 
Using these techniques, we can design a composite pulse sequence, 
which compensates for the pulse length error up to an arbitrary higher 
order in perturbative expansion.

For a target (\ref{target}) with angles $\theta$ and $\phi$, 
the time-symmetric BB1 consists of
\be
&&\theta_1=\theta_5=\theta/2,
\quad\ \,
\theta_2=\theta_3/2=\theta_4=\pi,
\nonumber\\
&&
\phi_1=\phi_5=\phi,
\qquad
\phi_2=\phi_4=\phi+\kappa,
\qquad
\phi_3=3\phi+\kappa,
\label{BB1}
\ee
with
\be
\kappa=\arccos[-\theta/(4\pi)].
\ee
Let us evaluate the dynamical phase associated with the time-symmetric BB1. 
First, we note from $U=\re^{-2\ri W^1}$ that
\be
V^1\ket{\psi_a}=\re^{-\ri W^1}\ket{\psi_a}=\pm\sqrt{\omega_a}\ket{\psi_a}.
\ee
Then, we have
\be
\gamma_{\rm d}^a(1)=-\bra{\psi_a}W^1\ket{\psi_a}=(-1)^{a+1}\theta_1\bm{n}_1\cdot\bm{n}/2=(-1)^{a+1}\theta/4.
\ee
Next we find from $\theta_2=\pi$ and $\phi_2=\phi+\kappa$ that
\begin{equation}
\gamma_{\rm d}^a(2)=-\bra{\psi_a}{V^1}^\dag W^2V^1\ket{\psi_a}=-\bra{\psi_a}W^2\ket{\psi_a}=
(-1)^{a+1}\pi\bm{n}_2\cdot\bm{n}/2
=(-1)^{a}\theta/8
\end{equation}
and
\be
\re^{-\ri W^2}\ket{\psi_a}=\ket{(-1)^a\bm{n}^\prime};
\qquad
\bm{n}^\prime=(\cos(\phi+2\kappa), \sin(\phi+2\kappa), 0).
\ee
This leads to
\be
\gamma_{\rm d}^a(3)=-\bra{\psi_a}{V^2}^\dag W^3V^2\ket{\psi_a}=(-1)^{a+1}\pi\ \bm{n}_3\cdot\bm
{n}^\prime=(-1)^{a}\theta/4.
\ee
By adding individual dynamical phases, we finally obtain
\be
\gamma_{\rm d}^a = 2\gamma_{\rm d}^a(1) +2 \gamma_{\rm d}^a(2)+\gamma_{\rm d}^a(3)=(-1)^{a+1}
(\theta/2-\theta/4-\theta/4)=0.
\ee
This result confirms that the time-symmetric BB1 is also a GQG.


\subsubsection{Knill\rq s sequence}
Knill\rq s sequence \cite{Ryan10,Souza11} is a $k=5$ time-symmetric composite pulse. 
This sequence implements the target $U$ given by
\be
\theta=\pi,
\qquad
\bm{n}=(\cos(\alpha-\pi/6), \sin(\alpha-\pi/6),0)
\ee
where $\alpha$ is a free parameter.
The sequence is 
defined by
\begin{equation}
\theta_i=\pi\ (1 \leq i \leq 5),
\qquad
\phi_1=\phi_5=\alpha+\pi/6,
\qquad
\phi_2=\phi_4=\alpha,
\qquad
\phi_3=\alpha+\pi/2.
\end{equation}
This sequence is used in experiments to maintain the coherence of nitrogen-vacancy centres in diamond \cite{Ryan10} and to decouple a system from the environment \cite{Souza11}. Note that this sequence is robust against not only the pulse 
length error, but also the off-resonance error \cite{Souza11}.

Let us calculate the dynamical phase. First, we have
\begin{equation}
\gamma_{\rm d}^a(1)
=-\bra{\psi_a}W^1\ket{\psi_a}
=(-1)^{a+1}\pi\bm{n}_1\cdot\bm{n}/2
=(-1)^{a+1}(\pi/2)\cos(\pi/3)
=(-1)^{a+1}\pi/4.
\end{equation}
We find $V^1\ket{\psi_a}=\ket{(-1)^a\bm{n}^\prime}$ with
\be
\bm{n}^\prime=(\cos(\alpha+\pi/2), \sin(\alpha+\pi/2),0).
\ee
Then, by the similar argument as that used for the first step, we have
\begin{equation}
\gamma_{\rm d}^a(2)
=0.
\end{equation}
Further, we observe $V^2\ket{\psi_a}=\ket{(-1)^a\bm{n}^{\prime\prime}}$ with
\be
\bm{n}^{\prime\prime}=(\cos(\alpha-\pi/2), \sin(\alpha-\pi/2),0).
\ee
Then, we have
\begin{equation}
\gamma_{\rm d}^a(3)
=(-1)^a\pi/2.
\end{equation}
We find, by adding individual dynamical phases,
\be
\gamma_{\rm d}^a = 2\gamma_{\rm d}^a(1) +2 \gamma_{\rm d}^a(2)+\gamma_{\rm d}^a(3)=(-1)^{a+1}
(\pi/2+0-\pi/2)=0.
\ee

This example shows that the composite pulses robust against several 
systematic errors 
are also GQGs, if they compensates for at least the pulse length error. 
Thus, by construction, 
the composite pulses proposed in \cite{Ichikawa11, Alway07}, which are simultaneously 
robust against the above two errors, are also GQGs.

\subsection{Two-Qubit System}
Since our interest lies in the CNOT operation, we choose the target
\be
U=\re^{-\im \Omega X/2},
\ee
which is the entangling part in the CNOT gate.
The cyclic state $\ket{\psi_a}$ is an eigenstate of $X$ in Eq.~(\ref{xyz}). 
In the binary notation $a=2p+q$ where $p, q\in\{0,1\}$, 
we find
\be
\ket{\psi_a}=\ket{p}\ot\ket{q}.
\ee


Jones designed a composite pulse sequence for a 
two-qubit system from a one-qubit
composite pulse sequence~\cite{Jones03}, by employing the
isomorphism among the generators given in 
Sec.~\ref{CPGQG}~(\ref{2qubit}). Let us introduce a notation 
$(\Omega)_\phi = \exp[-\ri\Omega(\cos\phi X + \sin\phi Y)/2]$ and set
the target to $(\pi/2)_0$ in this notation.
Jones' sequence is given by
\be
 (\pi/4)_0 (\pi)_\kappa (2\pi)_{3\kappa} (\pi)_\kappa (\pi/4)_0,
 \qquad
 \kappa=\arccos(-1/8).
 \label{jones_seq}
\ee
Since the isomorphism maps $X$, $Y$, and $Z$ to
the Pauli matrices $\sigma_x$, $\sigma_y$, and $\sigma_z$, respectively,
Jones' sequence is a two-qubit analogue of
the BB1 sequence: the combination of the first and last pulses is the target pulse ($\theta=\pi/2$, $
\phi=0$)
and the others are the same as the BB1 sequence (\ref{BB1}). Similarly, 
the composite pulses in \cite
{Hill07, Testolin07, Tomita10} are the two-qubit counterparts of those 
in \cite{Brown04}.

Evaluation of the dynamical phase is easy if we  make use of the 
isomorphism already mentioned. Since
$X$ is mapped to $\sigma_x$, the cyclic vector $\ket{p}\ot\ket{q}$ should be 
sent to $\ket{(-1)^{p+q} \hat{\bm{x}}}$, which is also an eigenvector of the 
target $U=\exp(-\ri\pi\sigma_x/4)$. Thus 
the dynamical phase of Jones' sequence is transferred to that of the BB1 
sequence, which leads to
\be
\gamma_{\rm d}^{a}=0,
\ee
showing the sequence has vanishing dynamical phase. One can also achieve 
the same result by direct calculation without employing the isomorphism.

\section{Two Composite ${\bm{z}}$-Rotations}

In NMR, rotations around the $z$-axis must be implemented by a sequence of pulses,
since the rf-pulses (\ref{decW1}) have the restriction $\bm{n}_i\perp\hat{\bm{z}}$.
Thus, it is of interest to investigate whether the sequences are geometric.

First, we consider the following $k=3$ sequence to realise a target $U=\re^{-\ri\theta\sigma_z/2}$:
\be
\theta_1=\theta_3=\pi/2,
\qquad
\theta_2=\theta,
\qquad
\phi_1=-\phi_3=\pi/2,
\qquad
\phi_2=0.
\label{cpz1}
\ee
The cyclic states are $\ket{\psi_a}=\ket{(-1)^a\hat{\bm{z}}}=\ket{a}$. Let us calculate the dynamical phase.
The first one is
\be
\gamma_{\rm d}^a(1)=-(\pi/4)\bra{\psi_a}\hat{\bm{y}}\cdot\bm{\sigma}\ket{\psi_a}=(-1)^{a+1}(\pi/4)\hat{\bm{y}}\cdot\hat{\bm z}=0.
\ee
We find $V^1\ket{\psi_a}=\ket{(-1)^a\hat{\bm x}}$, which leads to
\be
\gamma_{\rm d}^a(2)=-(\theta/2)\bra{(-1)^a\hat{\bm x}}\hat{\bm{x}}\cdot\bm{\sigma}\ket{(-1)^a\hat{\bm x}}=(-1)^{a+1}\theta/2.
\ee
Furthermore, we obtain $V^2\ket{\psi_a}=\exp[{(-1)^{a+1}\ri\theta/2]}\ket{(-1)^a\hat{\bm x}}$.
Thus, we observe
\be
\gamma_{\rm d}^a(3)=(\pi/4)\bra{(-1)^a\hat{\bm x}}\hat{\bm{y}}\cdot\bm{\sigma}\ket{(-1)^a\hat{\bm x}}=(-1)^{a}(\pi/4)\hat{\bm{y}}\cdot\hat{\bm x}=0.
\ee
We conclude
\be
\gamma_{\rm d}^a=(-1)^{a+1}\theta/2\neq0.
\ee
Hence the pulse sequence (\ref{cpz1}) is not a GQG. Note that this sequence is not robust against the pulse length error, that is, $\Delta W\neq0$, which is exactly the contraposition of our claim.

\begin{figure}[t]
\begin{center}
  (a)\includegraphics[width=2.2in]{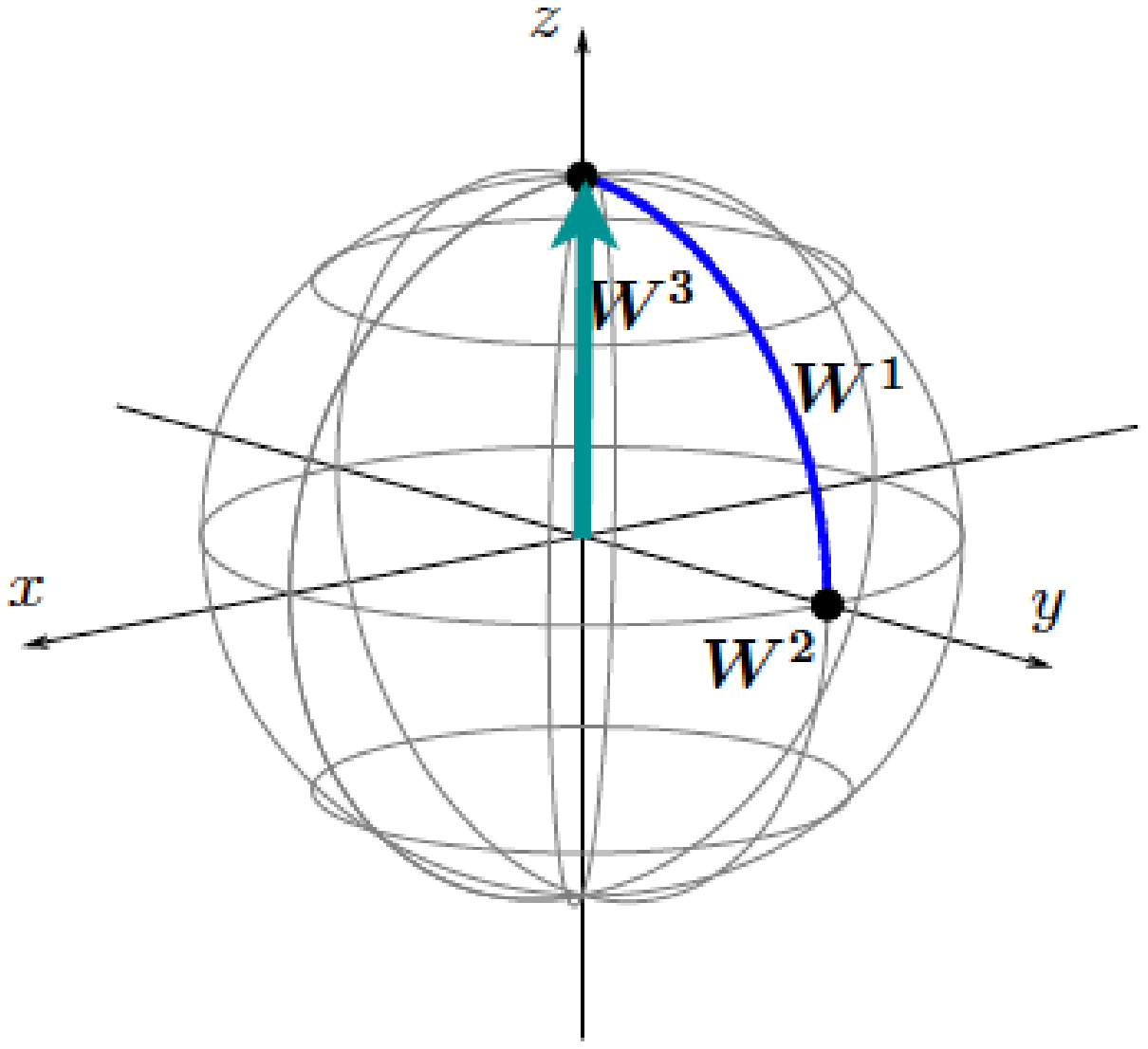}
(b)    \includegraphics[width=2.2in]{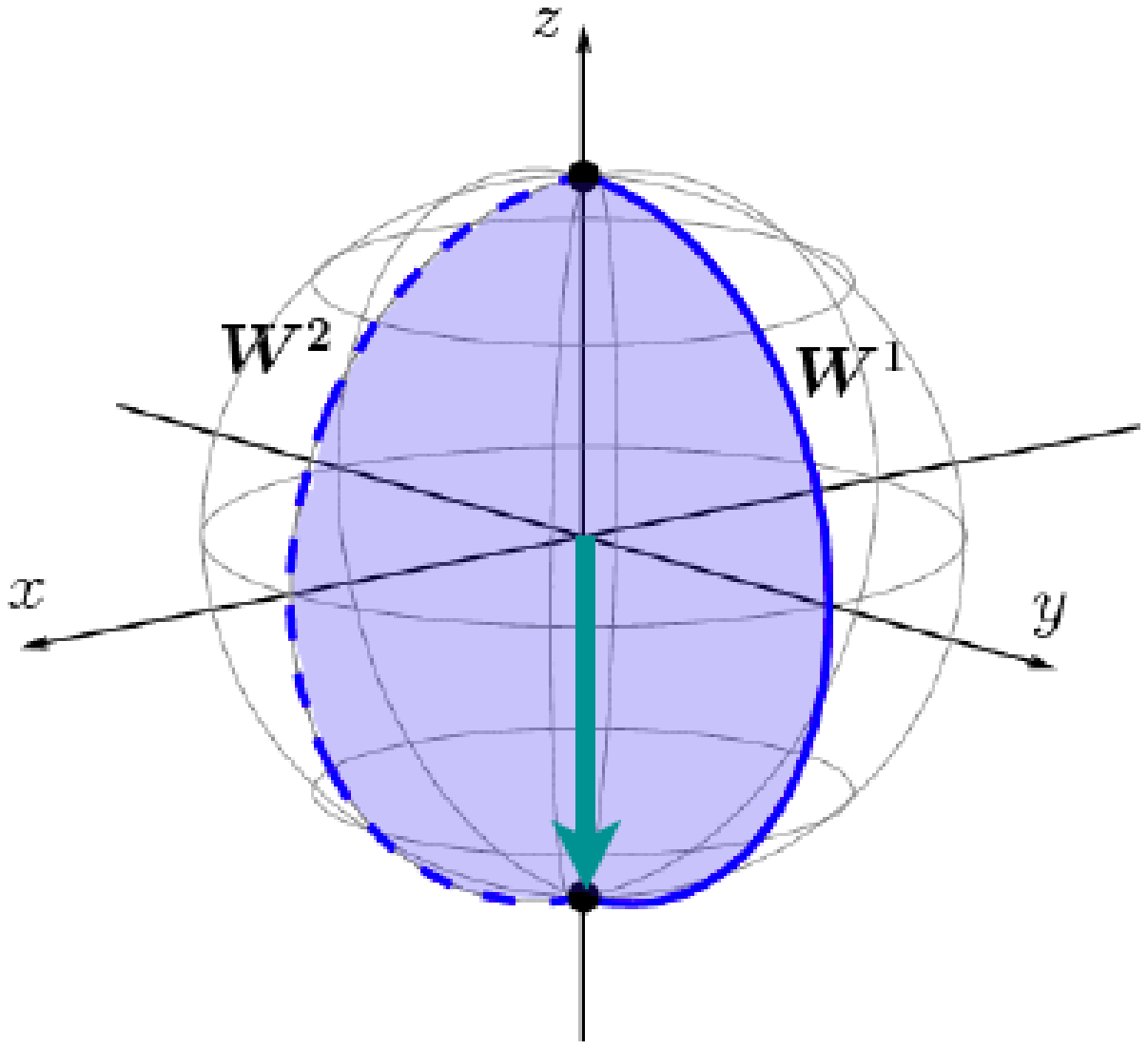}
\caption{(Colour online) Excursions of the cyclic states on the Bloch sphere. 
(a) The trajectory of the cyclic state $\ket{\hat{\bm z}}$ under the pulse sequence (\ref{cpz1}). 
Note that the trajectory fails to close, which shows that this sequence is dynamical.
(b) The trajectory of the cyclic state $\ket{-\hat{\bm z}}$ under the pulse sequence (\ref{cpz2}) for $\theta=\pi$. The solid angle subtended by the trajectory of the Bloch vector is $\pi = \theta$, which shows the geometric nature of the sequence.}
\label{CZs}
\end{center}
\end{figure}

Second, we investigate a $k=2$ pulse for $U=\re^{\ri\theta\sigma_z/2}$:
\be
\theta_1=\theta_2=\pi,
\qquad
\phi_1=0,
\qquad
\phi_2=\theta/2.
\label{cpz2}
\ee
The cyclic states are the same as those of the previous sequence.
We have
\be
\gamma_{\rm d}^a(1)=-(\pi/2)\bra{\psi_a}\hat{\bm{x}}\cdot\bm{\sigma}\ket{\psi_a}=(-1)^{a+1}(\pi/2)\hat{\bm{x}}\cdot\hat{\bm z}=0.
\ee
By the same way, we compute
\be
\gamma_{\rm d}^a(2)=0,
\ee
which clearly shows
\be
\gamma_{\rm d}^a=0.
\ee
Hence the pulse sequence (\ref{cpz2}) is a GQG.
This pulse is not robust against the pulse length error.
Indeed, we may check
\be
\Delta W=\epsilon\frac{\pi}{2}\[\(1+\cos\frac{\theta}{2}\)\sigma_x-\sin\frac{\theta}{2}\sigma_y\]\neq0
\ee
by direct calculation. This also tells us that not all GQGs are robust against the pulse length error.
The difference of these two composite $z$-rotations are visualised in Fig.~\ref{CZs}.

\section{Conclusion and Discussions}
In this article, we uncovered the relation between GQGs and the composite 
pulses robust against certain kinds of systematic errors. For the 
error (\ref{eW}), proportional to the Hamiltonian times the operation time, 
the compensation of the error automatically leads to vanishing dynamical 
phase. Thus, a non-trivial operation by a composite pulse robust against such
an error is a GQG.

We pointed out that there are two kinds of errors assuming the form (\ref{eW}).
One is the pulse length error and the other is the $J$-coupling error. 
This implies that the composite pulses robust against these errors are GQGs. 
This observation was illustrated and confirmed by directly showing that
the dynamical phase vanishes for several typical composite pulses: $90^\circ
$-$180^\circ$-$90^\circ$, SCROFULOUS, BB1, Knill\rq s sequence for 
the pulse length error and Jones' pulse sequence for the $J$-coupling error. 
The two most common composite $z$-rotations were also examined.

Our work has shown that we can construct a universal gate set composed of 
GQGs simply by using the composite pulses.
This suggests that NMR is quite a useful test bench of geometric quantum
computation. In view of this, further study of composite pulses, 
{\it e.g.} \cite{Ota09a}, is desirable for deeper understanding of the 
geometric quantum computation.

\begin{acknowledgements}
We would like to thank Jonathan~A.~Jones and Yukihiro~Ota for valuable discussions and
Dieter~Suter for drawing our attention to 
Knill\rq s sequence.
This work is supported by `Open Research Center' Project for Private Universities: matching fund subsidy from MEXT (Ministry of Education, Culture, Sports, Science and Technology), Japan. YK and MN would like to thank partial supports of Grants-in-Aid for 
Scientific Research from the JSPS (Grant No.~23540470).
\end{acknowledgements}

\appendix{Geometry of Aharonov-Anandan Phase}
\label{appHol}

In this appendix, we outline the relevant aspects of the Aharonov-Anandan
phase in the context of the present article. The geometric nature of the
Aharonov-Anandan phase is derived from that of the fibre bundle structure
associated with the Hilbert space. See \cite{Bengtsson06, Nakahara03}
for technical details.

\begin{figure}[t]
\begin{center}
  \includegraphics[width=2.5in]{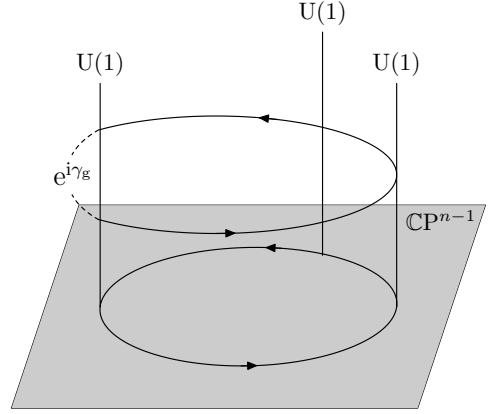}
\caption{Schematic diagram of Aharonov-Anandan phase. The set of the normalised states forms 
${S}^{2n-1}\subset\C^n$ and this subset ${S}^{2n-1}$ can be seen as a U(1)-bundle 
over the projective Hilbert space $\C P^{n-1}$. Given a closed path in the base manifold $\C P^{n-1}$, the horizontal lift of the path is naturally defined by a connection in the U(1)-bundle. 
The holonomy associated with the horizontal lift is given 
as $\re^{\ri\gamma_{\rm g}}\in$U(1), which can be seen as a global phase difference accumulated 
through the parallel transport along the horizontal lift on the path in $\C P^{n-1}$.}
\label{traverse}
\end{center}
\end{figure}

Consider the Hilbert space $\C^n$. In quantum mechanics, we are
exclusively concerned with the set of normalised vectors in $\C^n$.
The set of normalised vectors form the $(2n-1)$-dimensional sphere 
${S}^{2n-1} \subset\C^n$. 
Moreover, we need to identify vectors that differ by an overall phase;
two normalised states
$\ket{\psi}$ and $\re^{\ri\gamma}\ket{\psi}$ represent the identical physical
state for any $\gamma \in \R$. The manifold obtained from $S^{2n-1}$
under this identification is called the complex projective space;
$$
\C{P}^{n-1} \simeq S^{2n-1}/{\rm U}(1),
$$
where U(1) is the set of overall phases.

For $n=2$, we obtain $\C{P}^1=S^2$, which is nothing but the Bloch sphere.
Accordingly, $S^3$ is identified with a U(1)-bundle over $S^2$ (the Hopf 
fibration).
More generally, $S^{2n-1}$ is a U(1)-bundle over the base manifold $\C{P}^{n-1}$.
A point in $\C{P}^{n-1}$ represents a physical state and its phase freedom
is represented by the fibre U(1). The identification naturally introduces
the projection $\pi: S^{2n-1} \to \C{P}^{n-1}$. Fixing the phase is
equivalent to taking a point in the fibre (See Fig. \ref{traverse}).

It should be noted that $\C P^{n-1}$ has a natural metric called the
Fubini-Study metric. Given a metric in the base manifold, we can construct a connection in the base manifold. This defines the horizontal lift of a given curve in the base manifold to the fibre bundle $S^{2n-1}$. Now suppose that there is a closed loop in the base manifold. If one carries a point on
a fibre over $p \in \C{P}^{n-1}$ along the horizontal lift of the loop, the
point comes back to a point in the same fibre, which is not necessary the 
initial point. 
This U(1) phase factor obtained after traversing a loop is called the holonomy
associated with the loop and the horizontal lift. The Aharonov-Anandan phase
is nothing but this U(1) phase factor, which 
is geometric in the sense that it depends only on the loop in the base manifold and the connection of the U(1)-bundle but not
on how fast the loop is traversed. 

We note that the twice the Aharonov-Anandan phase is the solid angle at the 
origin subtended by the trajectory of a state vector on the Bloch sphere 
($\C P^1$, in this case) during a 1-qubit operation.

\appendix{Proof of Eq.~(\ref{gg})}
\label{appA}

In this appendix we prove Eq.~(\ref{gg}).
For this purpose, we first note the identity
\be
\sigma_z W^i\sigma_z=-W^i
\label{zWz}
\ee
for $W^i$ of Eq.~(\ref{decW1}), because $\bm{n}_i \perp \hat{\bm{z}}$.
Multiplying $-\ri$ and exponentiating Eq.~(\ref{zWz}), we find
\be
\sigma_z\re^{-\ri W^i}\sigma_z=\re^{\ri W^i}.
\ee
Then, we obtain
\be
{V}^{k-i}=\re^{\ri W^i}\cdots\re^{\ri W^1}U=\sigma_zV^i\sigma_zU
\label{cor}
\ee
for a time-symmetric composite pulse. It then follows that
\be
\ket{\psi_a(k-i)}
&=&V^{k-i}\ket{\psi_a}\nonumber\\
&=&\sigma_zV^i\sigma_zU\ket{\psi_a}\qquad\text{ from Eq.~(\ref{cor})}\nonumber\\
&=&\omega_a\sigma_zV^i\sigma_z\ket{\psi_a}\qquad \text{from Eq.~(\ref{eigenU})}\nonumber\\
&=&\omega_a\sigma_zV^i\ket{\psi_{a\oplus1}}\nonumber\\
&=&\omega_a\sigma_z\ket{\psi_{a\oplus1}(i)},
\label{cor2}
\ee
where we denote the sum modulo two by $\oplus$. Therefore, using the
condition
$\Tr\ W^i=0$ and the completeness relation with respect to
$\{\ket{\psi_a(i)}\}_{a=1,2}$, we observe that
\be
\gamma_{\rm d}^a(k+1-i)&=&-\bra{\psi_a(k-i)}W^{k+1-i}\ket{\psi_a(k-i)}\nonumber\\
&=&-\bra{\psi_{a\op1}(i)}\sigma_zW^i\sigma_z\ket{\psi_{a\op1}(i)}\qquad\text{from Eq.~(\ref{cor2}) and $|\omega_a|^2=1$}\nonumber\\
&=&\bra{\psi_{a\op1}(i)}W^i\ket{\psi_{a\op1}(i)}\qquad\qquad\quad \text{from Eq.~(\ref{zWz})}\nonumber\\
&=&\Tr\[W^i\(\mathbbm{1}_2-\ket{\psi_a(i)}\bra{\psi_a(i)}\)\]\nonumber\\
&=&-\bra{\psi_{a}(i)}W^i\ket{\psi_{a}(i)} \qquad\qquad\qquad \text{from $\Tr W^i=0$}\nonumber\\
&=&-\bra{\psi_{a}(i-1)}W^i\ket{\psi_{a}(i-1)}\nonumber\\
&=&\gamma_{\rm d}^{a}(i),
\ee
which proves Eq.~(\ref{gg}).



\begin{thebibliography}{99}
%
\bibitem{Freeman99}
Freeman, R. 1999
{\it Spin Choreography}.
Oxford: Oxford University Press. 
%
\bibitem{Claridge99}
Claridge, T. D. W. 1999
{\it High-Resolution NMR Techniques in Organic Chemistry}. 
Oxford: Elsevier.
%
\bibitem{Jones09}
Jones, J. A. 2009
Composite pulses in NMR quantum computation.
{\it J. Ind. Inst. Sci.} {\bf 89}, 303-308.
%
\bibitem{Levitt79}
Levitt, M. H. \& Freeman, R. 1979
NMR Population Inversion Using a Composite Pulse.
{\it J.~Magn.~Reson.} {\bf 33}, 473-476.
(doi: 10.1016/0022-2364(79)90265-8)
%
\bibitem{Levitt86}
Levitt, M. H. 1986
Composite Pulses.
{\it Prog. NMR Spectrosc.} {\bf 18}, 61-122.
(doi: 10.1016/0079-6565(86)80005-X)
%
\bibitem{Levitt96}
Levitt, M. H. 1996
Composite Pulses.
In {\it Encyclopedia of Nuclear Magnetic Resonance}
(eds. D. M. Grant and R. K. Harris),
pp. 1396-1411. 
Sussex: Wiley.
%
\bibitem{Tycko85}
Tycko, R., Pines, A. \& Guckenheimer J. 1985
Fixed point theory of iterative excitation schemes in NMR.
{\it J. Chem. Phys.} {\bf 83}, 2775-2802. 
(doi: 10.1063/1.449228)
%
\bibitem{Khaneja05}
Khaneja, N., Reiss, T., Kehlet, C., Schulte-Herbr\"uggen, T. \& Glaser, S. J. 2005
Optimal control of coupled spin dynamics: design of NMR pulse sequences by gradient ascent algorithms.
{\it J. Magn. Reson.} {\bf 172}, 296-305. 
(doi: 10.1016/j.jmr.2004.11.004)
%
\bibitem{Machnes11}
Machnes, S., Sander, U., Glaser, S. J., de Fouqui\`eres P.,  Gruslys, A., Schirmer, S. \& Schulte-Herbr\"uggen, T. 2011
Comparing, optimizing, and benchmarking quantum-control algorithms in a unifying programming framework.
{\it Phys. Rev. A} {\bf 84}, 022305.
(doi: 10.1103/PhysRevA.84.022305)
%
\bibitem{Ichikawa11}
Ichikawa, T., Bando,~M., Kondo, Y. \& Nakahara, M. 2011
Designing Robust Unitary Gates: Application to Concatenated Composite Pulse.
{\it Phys. Rev. A} {\bf 84}, 062311.
(doi: 10.1103/PhysRevA.84.062311)
%
\bibitem{Barenco}
Barenco, A., Bennett, C. H., Cleve, R., DiVincenzo, D. P., Margolus, N., Shor, P., Sleator, T., Smolin, J. A. \& Weinfurter, H. 1995 Elementary gates for quantum computation.
{\it Phys. Rev. A} {\bf 52}, 3457-3467.
(doi: 10.1103/PhysRevA.52.3457)
%
\bibitem{Nielsen00}
Nielsen, M. A. \& Chuang, I. C. 2000
{\it Quantum Information and Quantum Computation}.
Cambridge: Cambridge University Press.
%
\bibitem{Bengtsson06}
Bengtsson, I. \& \.Zyczkowski, K. 2006
{\it Geometry of Quantum States},
New York: Cambridge University Press.
%
\bibitem{Gaitan08}
Gaitan, F. 2008
{\it Quantum Error Correction and Fault Tolerant Quantum Computing}.
Boca Raton: Taylor \& Francis.
%
\bibitem{Nakahara08}
Nakahara, M. \& Ohmi, T. 2008
{\it Quantum Computing: From Linear Algebra to Physical Realizations}.
Boca Raton: Taylor \& Francis.
%
\bibitem{Jones11}
Jones, J. A. 2011
Quantum Computing with NMR.
{\it Prog. NMR Spectrosc.} {\bf 59}, 91-120.
(doi: 10.1016/j.pnmrs.2010.11.001)
%
\bibitem{Jones03}
Jones, J. A. 2003
Robust Ising Gates for Practical Quantum Computation.
{\it Phys. Rev. A.} {\bf 67}, 012317.
(doi: 10.1103/PhysRevA.67.012317)
%
\bibitem{Hill07}
Hill, C. D. 2007
{\rm Robust Controlled-NOT Gates from Almost Any Interaction}.
{\it Phys. Rev. Lett.} {\bf 98}, 180501.
(doi: 10.1103/PhysRevLett.98.180501)
%
\bibitem{Testolin07}
Testolin, M. J., Hill, C. D., Wellard, C. J. \& Hollenberg, L. C. L. 2007
Robust controlled-NOT gate in the presence of large fabrication-induced variations of the exchange interaction strength.
{\it Phys. Rev. A} {\bf 76}, 012302.
(doi: 10.1103/PhysRevA.76.012302)
%
\bibitem{Tomita10}
Tomita, Y., Merrill, J. T. \& Brown, K. R. 2010
{\rm Multi-Qubit Compensation Sequences}.
{\it New J. Phys.} {\bf 12}, 015002.
(doi: 10.1088/1367-2630/12/1/015002)
%
\bibitem{Cummins03}
Cummins, H. K., Llewellyn, G. \& Jones, J. A. 2003
Tackling systematic errors in quantum logic gates with composite rotations.
{\it Phys. Rev. A} {\bf 67}, 042308.
(doi: 10.1103/PhysRevA.67.042308)
%
\bibitem{Brown04}
Brown, K. R., Harrow, A. W., \& Chuang, I. L. 2004
Arbitrarily accurate composite pulse sequences.
{\it Phys. Rev. A} {\bf 70}, 052318.
(doi: 10.1103/PhysRevA.70.052318)
%
\bibitem{Alway07}
Alway, W. G. \& Jones, J. A. 2007
Arbitrary precision composite pulses for NMR quantum computing.
{\it J. Magn. Reson.} {\bf 189}, 114-120.
(doi: 10.1016/j.jmr.2007.09.001)
%
\bibitem{Zanardi99}
Zanardi, P. \& Rasetti, M. 1999
{\rm Holonomic Quantum Computation}.
{\it Phys. Lett. A} {\bf 264}, 94-99.
(doi: 10.1016/S0375-9601(99)00803-8)
%
\bibitem{Zhu00}
Zhu, S.-L., and Wang, Z.-D., 2002
Implementation of universal quantum gates based on nonadiabatic geometric phases,
{\it Phys. Rev. Lett.} {\bf 89} 097902.
(doi: 10.1103/PhysRevLett.89.097902)
%
\bibitem{Shapere89}
Shapere,~A. \& Wilczek,~F. 1989
{\it Geometric phases in physics},
Singapore: World Scientific.
%
\bibitem{Berry84}
Berry,~M.~V. 1984
Quantal Phase Factors Accompanying Adiabatic Changes.
{\it Proc. R. Soc.  A} {\bf 392}, 45-57.
(doi: 10.1098/rspa.1984.0023)
%
\bibitem{Wilczek84}
Wilczek, F. \& Zee, A. 1984
Appearance of Gauge Structure in Simple Dynamical Systems.
{\it Phys. Rev. Lett.} {\bf 52}, 2111-2114.
(doi: 10.1103/PhysRevLett.52.2111)
%
\bibitem{Aharonov87}
Aharonov, Y. \& Anandan, J. 1987
{\rm Phase Change during a Cyclic Quantum Evolution}.
{\it Phys. Rev. Lett.} {\bf 58}, 1593-1596.
(doi: 10.1103/PhysRevLett.58.1593)
%
\bibitem{Mead92}
Mead C. A. 1992
{\rm The geometric phase in molecular systems}.
{\it Rev. Mod. Phys.} {\bf 64}, (1992) 51-85.
(doi: 10.1103/RevModPhys.64.51)
%
\bibitem{Simon83}
Simon, B. 1983
Holonomy, the Quantum Adiabatic Theorem, and Berry's Phase,
{\it Phys. Rev. Lett.} {\bf 51}, 2167-2170.
(doi: 10.1103/PhysRevLett.51.2167)
%
\bibitem{Nakahara03}
Nakahara, M. 2003
{\it Geometry, Topology and Physics}, 2nd edn,
Boca Raton: Taylor \& Francis.
%
\bibitem{Page87}
Page, D. N. 1987
{\rm Geometrical description of Berry\rq s phase}.
{\it Phys. Rev. A} {\bf 36}, 3479-3481.
(doi: 10.1103/PhysRevA.36.3479)
%
\bibitem{Zhu05}
Zhu, S.-L. \& Zanardi, P. 2005
Geometric quantum gates that are robust against stochastic control errors.
{\it Phys. Rev. A} {\bf 72}, 020301.
(doi: 10.1103/PhysRevA.72.020301)
%
\bibitem{Blais03}
Blais, A. \& Tremblay, A.-M. S. 2003
Effect of noise on geometric logic gates for quantum computation.
{\it Phys. Rev. A} {\bf 67}, 012308.
(doi: 10.1103/PhysRevA.67.012308)
%
\bibitem{Kondo11}
Kondo,~Y. \& Bando,~M. 2011
Geometric Quantum Gates, Composite Pulses, and Trotter-Suzuki Formulas.
{\it J. Phys. Soc. Jpn.} {\bf 80}, 054002. 
(doi: 10.1143/JPSJ.80.054002)
%
\bibitem{Ota09a}
Ota, Y. \& Kondo, Y. 2009
Composite pulses in NMR as nonadiabatic geometric quantum gates.
{\it Phys. Rev. A} {\bf 80}, 024302.
(doi: 10.1103/PhysRevA.80.024302)
%
\bibitem{Ota09b}
Ota, Y., Goto, Y., Kondo, Y. \& Nakahara, M. 2009
{\rm Geometric quantum gates in liquid-state NMR based on a cancellation of dynamical phases}.
{\it Phys. Rev. A} {\bf 80}, 052311.
(doi: 10.1103/PhysRevA.80.052311)
%
\bibitem{Wimperis94}
Wimperis, S. 1994
Broadband, Narrowband, and Passband Composite Pulses for Use in Advanced NMR Experiments.
{\it J. Magn. Reson. A} {\bf 109}, 221-231.
(doi: 10.1006/jmra.1994.1159)
%
\bibitem{Ryan10}
Ryan, C. A., Hodges, J. S. \& Cory, D. G. 2010
Robust Decoupling Techniques to Extend Quantum Coherence in Diamond.
{\it Phys. Rev. Lett.} {\bf 105}, 200402.
(doi: 10.1103/PhysRevLett.105.200402)
%
\bibitem{Souza11}
Souza, A. M., \'Alvarez G. A. \& Suter, D. 2011
Robust Dynamical Decoupling for Quantum Computing and Quantum Memory.
{\it Phys. Rev. Lett.} {\bf 106}, 240501.
(doi: 10.1103/PhysRevLett.106.240501)
%
\end{thebibliography}
\end{document}